# Untangling the Furball: A Practice Mapping Approach to the Analysis of Multimodal Interactions in Social Networks

AXEL BRUNS, KATERYNA KASIANENKO, VISHNU PADINJAREDATH SURESH, EHSAN DEHGHAN, LAURA VODDEN

DIGITAL MEDIA RESEARCH CENTRE, QUEENSLAND UNIVERSITY OF TECHNOLOGY

BRISBANE, AUSTRALIA

A.BRUNS@QUT.EDU.AU, KATERYNA.KASIANENKO@HDR.QUT.EDU.AU, VISHNU.PRASAD@HDR.QUT.EDU.AU, E.DEHGHAN@QUT.EDU.AU, LAURA.VODDEN@QUT.EDU.AU

## Abstract

This article introduces the analytical approach of *practice mapping*, using vector embeddings of network actions and interactions to map commonalities and disjunctures in the practices of social media users, as a framework for methodological advancement beyond the limitations of conventional network analysis and visualisation. In particular, the methodological framework we outline here has the potential to incorporate multiple distinct modes of interaction into a single practice map, can be further enriched with account-level attributes such as information gleaned from textual analysis, profile information, available demographic details, and other features, and can be applied even to a cross-platform analysis of communicative patterns and practices.

The article presents practice mapping as an analytical framework and outlines its key methodological considerations. Given its prominence in past social media research, we draw on examples and data from the platform formerly known as Twitter in order to enable experienced scholars to translate their approaches to a practice mapping paradigm more easily, but point out how data from other platforms may be used in equivalent ways in practice mapping studies. We illustrate the utility of the approach by applying it to a dataset where the application of conventional network analysis and visualisation approaches has produced few meaningful insights.

## Introduction

Social network analysis has become a key tool for the study of user actions and interactions on contemporary social media platforms, and beyond. Often, however, such analyses remain somewhat superficial, merely presenting the standard network graphs produced by the key visualisation algorithms implemented in popular tools like Gephi (Bastian et al., 2009), or identifying the clusters of tightly connected accounts highlighted by popular modularity algorithms like Louvain (Blondel et al., 2008), without sufficient consideration of the limitations of such approaches. At worst, and especially in the hands of inexperienced researchers, such attempts to make sense of networks result in network 'furballs' of severely limited value (and the authors of this article readily admit to having produced such visualisations ourselves at times); even if the network analysis and visualisation produces networks with more distinct features, however, they often fail to represent more than a handful of obvious patterns, reducing complex and multilayered action and interaction patterns into overly simplified graphs and statistics.

Fig. 1, for instance, shows a typical network graph of Twitter interactions (@mentions and retweets) from a controversial political debate. At face value, the network visualisation itself, as well as the interaction data underlying it, provide very limited insight into the structures and dynamics of interaction within the participating group of accounts; while the modularity algorithm appears to have identified a number of distinct network clusters, their validity must be questioned since the visual network structure does not provide any strong evidence that supports their existence. It is possible to derive meaningful insights from such data, however, as we will show in the following.

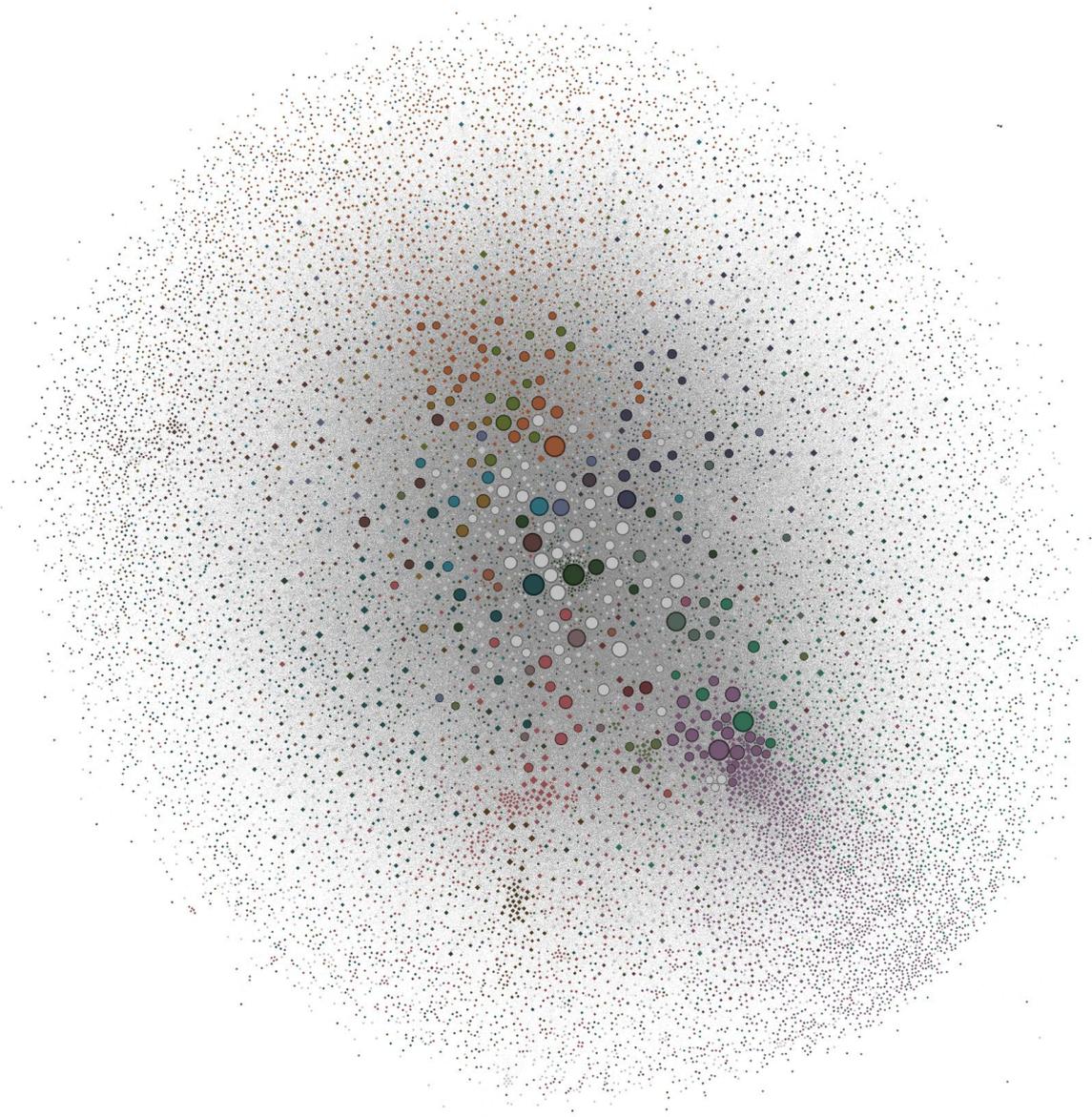

Fig. 1: A typical 'furball' network of @mentions and retweets on a controversial topic on Twitter.

A number of key factors combine to produce such pragmatic but undesirable simplifications. First, in the context of social media platforms (as well as in many other comparable communicative situations), user actions and interactions are often multimodal: on Twitter, for instance, users were able to @mention, retweet, quote-tweet, follow, and direct-message each other, but in many publications drawing on Twitter data some or all of these interactions were combined to produce a single connection (i.e. network edge) between two accounts, whose strength (i.e. edge weight) represented the total number of @mentions, retweets, quote-tweets, and/or other forms of interaction from one user to another. Alternatively, some publications resorted to presenting multiple separate graphs of the different interaction types – Williams et al. (2015), for instance, show separate graphs of @mentions, retweets, and follower-followee relationships for each of the three Twitter hashtags they examined. As Howison et al. (2011) argue, the conceptual choice behind selecting one type of interaction over others or combining different types of interactions is rarely theorised, which affects the interpretability of such studies. Extra dimensions are often simplified into single edges between the nodes, which eliminates crucial distinctions in communication practices from the overall network.

Second, whether publications distinguish between different interaction types in this way or combined them into a single network edge to represent the connection between two accounts, such analyses also often struggle to fully consider (and visualise) directionality. Most social media interactions are not inherently reciprocal: while friendship relations between Facebook accounts are mutual, for instance, the same is not true for follower relations on Twitter, and communicative interactions between accounts (@mentions or retweets on Twitter, comments on Facebook or Reddit, etc.) are generally reciprocal only if the mentioned account chooses to respond. Standard network analysis tools like Gephi do provide the means to take the directionality of network edges into account in both analysis and visualisation – but such details are often lost in the published research. (Gephi, for instance, uses curved edges to embed a clockwise connection logic, but even articles that draw on this functionality rarely explain how the curvature of an edge indicates which is its source and which is its target node.)

Third, even if the directionality of the underlying network has been reflected in the network data, common implementations of network modularity algorithms (colloquially known as 'community detection' algorithms, though of course the clusters they detect may not represent true *communities* in the sense used by media, communication, and cultural studies scholars) must often necessarily ignore this directionality in order to produce meaningful results. The powerful Louvain algorithm (Blondel et al., 2008), for instance – available as a Gephi module and a stand-alone package for Python and other programming languages – does not take directionality into account as it identifies network clusters. This is problematic, since it will end up identifying, for instance, both the bots and the genuine users caught up in a mob of bot accounts trolling a number of target users through repeated @mentions as a tightly packed cluster (or even as a 'community', in the language Louvain itself uses), even if the targetted users at the centre of the mobbing remain entirely passive; or assign politicians, journalists, and ordinary users to a large cluster of accounts discussing politics if those ordinary users tweet frequently enough at the accounts of politicians and journalists, even though the latter two types of accounts engage only with each other and never respond to members of the general public.

We provide this brief list of key concerns about the limitations of conventional social network analysis and visualisation not as a fundamental critique of network analytics as a valuable component of the communication researcher's toolkit – we have used many of these techniques ourselves, and at times struggled with their limitations, yet value the contributions they can make if used with the necessary care and rigour. However, we also suggest that social network analysis – or, more to the point, *the analysis of participant practices in social networks* – need not remain constrained by these limitations. Rather, in this article we present the concept of *practice mapping*, and the use of vector embeddings of network actions and interactions as a means to map such practices, as a framework for methodological advancement beyond the limitations of conventional network analysis and visualisation. In particular, the methodological framework we outline here has the potential to incorporate multiple distinct modes of interaction into a single practice map, can be further enriched with account-level attributes such as information gleaned from textual analysis, profile information, available demographic details, and other features, and can be applied even to a cross-platform analysis of communicative patterns and practices.

This article, then, is designed to present practice mapping as an analytical framework and outline its key methodological considerations. Given its prominence in past social media research, we draw on examples and data from the platform formerly known as Twitter in order to enable experienced scholars to translate their approaches to a practice mapping paradigm more easily, but point out throughout how data from other platforms may be used in equivalent ways in practice mapping studies. We begin our discussion with an introduction to the concept of practice mapping, explaining why we have chosen this term; we then outline the key steps in its methodological implementation; this is followed by a brief example that illustrates its utility. We conclude with a discussion of the applicability and limitations of the practice mapping approach, and an outlook on further developments.

## Why 'Practice Mapping'?

As we have noted, conventional network analysis and visualisation maps unidirectional and/or mutual connections between two accounts that might represent multiple modes of interaction between them, and often flattens them into a single network representation. This often oversimplifies the structure of the network in order to be able to perform any network analysis at all.

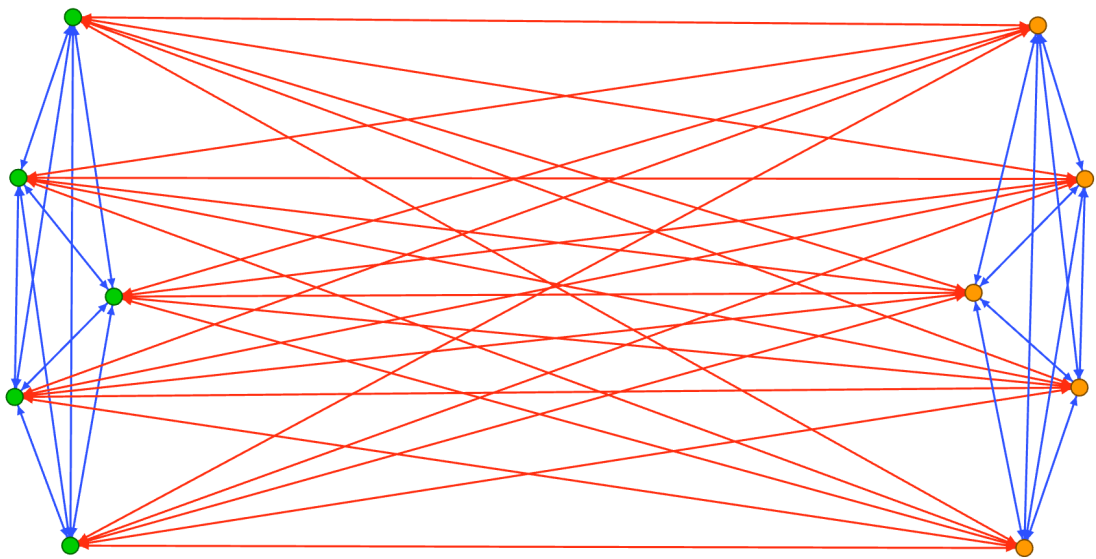

Fig. 2: Simplified representation of a highly polarised network – members of Group A (green, left) and Group B (orange, right) each only retweet within their own group (blue lines), and only @mention their opponents (red lines).

Consider the example shown in fig. 2: here, two opposing sides of a polarised debate (shown in green and orange, respectively) engage in in-group support by retweeting each other's posts (blue edges), as well as in out-group animosity by critically @mentioning members of the opposing side (red edges). A visualisation of this network in a standard network analysis tool like Gephi would identify members of both groups as belonging to the same densely connected network cluster because of the strength of their connections both within and across the two sides (fig. 3). This is correct from a purely functional point of view, but any interpretation of such a cluster as a 'community' in the colloquial as well as scholarly sense of the term – for example, as a community of practice (Lave & Wenger, 1991) – would be inherently incorrect.

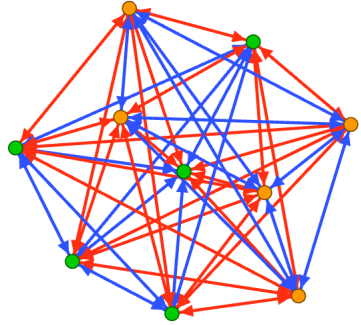

*Fig. 3: A Gephi visualisation of the network between Group A (green nodes) and Group B (orange nodes), using the Force Atlas 2 algorithm. Both groups exclusively engage in reciprocal in-group retweeting (blue lines between members of the same group) and out-group @mentioning (red lines between members of opposing groups).*

Instead of directly mapping the networks of connections between accounts, then, it would be preferrable instead to understand the activity and interactivity patterns of each account as its distinct participation *practice*, and to assess how similar this practice is to the distinct practices of all other accounts. Accounts in Group A that are frequently retweeting their friends and frequently @mentioning their opponents would thus be identified to have distinctly different practices from accounts in Group B that are frequently retweeting *their* friends and frequently @mentioning *their* opponents (fig. 4).

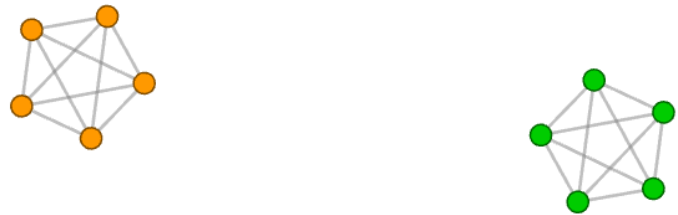

*Fig. 4: Gephi visualisation of the similarities in tweeting practices between members of the two groups, with grey lines representing the strength of similarity between the practices of any pair of two accounts.*

We describe this transformation of the underlying connection network into a network showing similarities between account actions and interactions as *practice mapping* because we can understand the sum total of each account's actions and interactions – including its patterns of engagement with other accounts, but also various other features from its use of language to its sharing of URLs, images, and videos – as its unique individual configuration of practices.

Scholars of sociology and media and communications understand 'practices' as patterns of 'doings and sayings' in a specific setting (Bakardjieva, 2020; Schatzki, 1996). Developed within fields of philosophy (Schatzki, 1996; Schatzki et al., 2000; Turner, 1994), education (Engeström, 2001), and organisational studies (Gherardi, 2000), the concept has been introduced to media and communication theory by Couldry (2004).

Couldry (2012) argued that the concept of practices is crucial for making sense of various phenomena ranging from everyday activities to political and social activism, as it allows studies to capture patterns of activity that display a level of regularity and sociality. There is no scholarly consensus on how often or how many times an action needs to be performed to be considered a practice. What is more important is for practices to be informed by specific knowledge and values, directed towards specific objects, and undertaken in ways that make relevant others recognise them as practices (Nicolini, 2012). For actions using the affordances of social media platforms, such recognition may be achieved by tapping into a particular 'texture of expression' (Papacharissi, 2015, p. 33). Drawing from Bourdieu's (1977, 1992) articulation of practice, Papacharissi (2015) sees practices on social media as emerging organically through processes of socialisation. Other social media scholars have increasingly adopted the notion of practice, though many have utilised the concept without deeper conceptualisation. While this is justifiable for a notion common to everyday language (Bakardjieva, 2020), this complicates both the empirical identification of practices and the further theoretical synthesis of the results of such studies.

There are some recent efforts to operationalise practice from a media and communication perspective. Most notably, Mattoni (2020) has proposed an operational definition of practices that positions them as phenomena more general than actions, performed by social actors through their use of skills in interaction with material objects and systems of meaning. Building on this definition, she developed a method of interviewing that produces 'media practice maps' which allowed her to explore political activists' daily practices of media use. To examine the practices of larger groups of users engaging with transnational movements like MeToo, Mendes and colleagues (2023) have enhanced the definition of platform vernaculars (Gibbs et al., 2015, p. 257) – or 'unique combinations of styles, grammars, and logics', specific to each social media platform within its geographic, social, and political contexts – to introduce a notion of 'vernacular practices' which can be identified through thematic analysis (Braun & Clarke, 2006). Building on the results of a cross-national interview study and systematic literature review, Trillò and co-authors (2022) produced a typology of social media rituals which they understand as 'typified communicative practices' (p. 1).

While these studies excel at capturing the social, symbolic, and material aspects of practices of smaller groups of users, qualitative approaches such as interviews or close reading are difficult to replicate at scale. This prevents us from understanding how practices are distributed within and across distinct groupings of users, how their prevalence changes with time, or how and whether they co-occur. Focussing on practices of knowledge-sharing on StackOverflow, Hillman and colleagues (2023) have demonstrated that 'trace ethnography' (Geiger & Ribes, 2011), or statistical analysis of large scale numerical and categorical data, complemented by interviews with high reputation scores and timeseries analysis of participants' data, can provide insights on how trajectories of knowledge-sharing change as participants become more experienced. Following Howison and colleagues' (2011, p. 790) signposting of practice theory as a fruitful theoretical grounding for network analysis, we argue that constructing a map of practices can offer a view of the connections, ruptures, and overlaps between the doings and sayings of much larger groups of users. A map of practices, constructed in the way we outline in the following, can also serve as a sampling frame for deeper contextual or interview analysis, which then adds the first-hand perspectives of practitioners to the birds-eye view offered by the map.

We note that the practice mapping approach we are outlining here is distinct from other well-known 'mapping' approaches such as issue mapping (Marres, 2015; Burgess & Matamoros-Fernández, 2016) or controversy mapping (Venturini & Munk, 2021), for two reasons. First, these approaches continue to draw on conventional analysis and visualisation of interaction networks between social media accounts, while practice mapping abstracts from these direct representations of communicative networks on social media platforms by mapping instead the patterns of similarity between the communicative practices of accounts in such networks. Second, issue and controversy mapping by definition focus on distinct topics and events, as identified for instance by a shared hashtag that is used by participating groups of accounts. While practice mapping can be applied to such datasets, too (as our discussion of the #robodebt example below will demonstrate), the existence of such a distinct issue or controversy is not required for its application: practice mapping can be used just as easily to identify distinct practices within a dataset of everyday activities within a given communicative context. The datasets our approach is applied to need not be determined by the existence of a thematic hashtag or other marker of a shared topic of discussion.

We also stress again that – in spite of our use of examples from Twitter as a convenient and well-understood platform example – the concept of practice mapping is by no means limited to this platform; indeed, we expect that the approach we outline here can be adapted for research tasks well outside the study of social networks in a narrow sense. As we have noted, an account's practices can include how it interacts with other accounts (in other words, depending on a platform's interactive affordances, how it views, clicks on, mentions, replies to, comments on, shares, likes, reacts to, up- and downvotes, mutes, blocks, reports, and otherwise responds to other accounts' posts); how it contributes original content of its own (e.g. what words, images, videos, URLs, hashtags, emoji, and other features it includes in its posts); where on the platform it does so (what pages, groups, hashtags, subreddits, discussion fora, group chats, channels, or other communicative spaces it posts to); what information about itself it or the platform provides (i.e. what profile details, activity metrics, or platform verification and participation badges are available for the account); or even when (during what hours of the day or days of the week) it is usually active. If available and ethically appropriate, such information might even be further enriched with other external information (for instance, party membership and roles for politicians' accounts; news organisation for journalists' accounts; etc.).

This does not imply that all such features are or should be treated as equivalent and exchangeable; how they might be weighted in an application of the practice mapping approach will depend on the specific questions to be explored by the research. While we cannot provide universally applicable guidelines for every possible use case, in the following section we discuss and illustrate how to operationalise the practice mapping framework for a given dataset, and how to consider the implications of weighting different communicative features.

## Practice Mapping in Practice

The central methodological innovation at the core of the practice mapping approach is to treat the interaction data that conventional network analysis would have used as a *direct* input, as well as any other activity patterns that can be derived from the data for each account, as input for an *intermediary* vector embedding stage of the data processing. Using the simple example of two polarised groups that we visualised in figs. 2-4, for instance, the first account in Group A – which we will name $A_1$ – engaged in the following directed interactions:

| Retweets | @mentions |
|---|---|
| $A_1 - 0$ | $A_1 - 0$ |
| $A_2 - 1$ | $A_2 - 0$ |
| $A_3 - 1$ | $A_3 - 0$ |
| $A_4 - 1$ | $A_4 - 0$ |
| $A_5 - 1$ | $A_5 - 0$ |
| $B_1 - 0$ | $B_1 - 1$ |
| $B_2 - 0$ | $B_2 - 1$ |
| $B_3 - 0$ | $B_3 - 1$ |
| $B_4 - 0$ | $B_4 - 1$ |
| $B_5 - 0$ | $B_5 - 1$ |

In other words, $A_1$ retweeted only the four other accounts in Group A (but not itself), and it @mentioned only the five accounts in Group B (but not any of the members of its own group). By treating every possible combination of account name and tweet type as a unique dimension, this means that we could represent $A_1$'s interaction practice as the 20-dimensional practice vector <0, 1, 1, 1, 1, 0, 0, 0, 0, 0, 0, 0, 0, 0, 0, 1, 1, 1, 1, 1>. If $A_2$ similarly retweeted each account in Group A but itself, and @mentioned every account in Group B, its vector would be <1, 0, 1, 1, 1, 0, 0, 0, 0, 0, 0, 0, 0, 0, 0, 1, 1, 1, 1, 1>, and so on. If accounts in Group B showed the opposite pattern, the comparison between vectors for $A_1$, $A_2$, $B_1$, and $B_2$ would look as follows:

|  | RT $A_1$ | RT $A_2$ | RT $A_3$ | RT $A_4$ | RT $A_5$ | RT $B_1$ | RT $B_2$ | RT $B_3$ | RT $B_4$ | RT $B_5$ | @$A_1$ | @$A_2$ | @$A_3$ | @$A_4$ | @$A_5$ | @$B_1$ | @$B_2$ | @$B_3$ | @$B_4$ | @$B_5$ |
|---|---|---|---|---|---|---|---|---|---|---|---|---|---|---|---|---|---|---|---|---|
| $A_1$ | 0 | 1 | 1 | 1 | 1 | 0 | 0 | 0 | 0 | 0 | 0 | 0 | 0 | 0 | 0 | 1 | 1 | 1 | 1 | 1 |
| $A_2$ | 1 | 0 | 1 | 1 | 1 | 0 | 0 | 0 | 0 | 0 | 0 | 0 | 0 | 0 | 0 | 1 | 1 | 1 | 1 | 1 |
| $B_1$ | 0 | 0 | 0 | 0 | 0 | 0 | 1 | 1 | 1 | 1 | 1 | 1 | 1 | 1 | 1 | 0 | 0 | 0 | 0 | 0 |
| $B_2$ | 0 | 0 | 0 | 0 | 0 | 1 | 0 | 1 | 1 | 1 | 1 | 1 | 1 | 1 | 1 | 0 | 0 | 0 | 0 | 0 |

Respectively, the pairs $A_1$ and $A_2$, and $B_1$ and $B_2$, each show considerably similar (but not entirely identical) practices, therefore, but there are no overlaps whatsoever between the practices of accounts in Group A and those of accounts in Group B.

In a second step, we can then apply a standard vector comparison approach to determine the similarity between the distinct practices of each pair of accounts. One common metric for doing so is cosine similarity, which is computed by taking the quotient of the dot-product of two vectors over the product of their magnitudes. In other words, cosine similarity parametrises the "pointing in the same direction"-ness of two vectors within a vector space as a comparative metric that produces real-valued outputs between 0 (not at all similar) and 1 (entirely identical). For the four sample accounts above, this results in cosine similarities of 0.89 between each pair of accounts in Group A and each pair of accounts in Group B, and cosine similarities of 0 for each pairing of an account of Group A with an account from Group B. (These cosine similarities never reach the maximum value of 1 because – at least in our theoretical example – none of the accounts ever retweets itself, so that retweet patterns even amongst accounts in the same group are never *entirely* identical.)

It is important to note here that these comparisons between the practice vectors for each pair of accounts are symmetrical: that is, the comparison of the vector for $A_1$ with the vector for $A_2$ produces the same value as the reverse comparison of $A_2$ with $A_1$. This halves the number of computational comparisons that need to be performed to compute these similarities; it also means that we can now analyse and visualise the set of pairwise comparisons between all accounts' practice vectors as an *undirected* network. For our simple thought example, a visualisation of the practice mapping network – where each account is a distinct network node, and the similarities between each pair of accounts' practices are represented by an edge of an appropriate weight between 0 and 1 – would result in a network map identical to that presented in fig. 4 above.

The example we have used here, however, represents an unusual case where all accounts directed exactly *one* retweet and/or @mention towards those accounts they chose to interact with. This is highly unlikely to occur in everyday practice. Taking a practice mapping approach, then, how should we treat cases where the two accounts we intend to compare show similar *overall* patterns in their engagement with others in the network, but have different levels of activity? For instance, if $A_1$ retweets $B_1$ fifty times and @mentions $B_2$ twenty times, and if $A_2$ retweets $B_1$ five times and @mentions $B_2$ two times, the overall distribution of their interactions across the network is identical, yet the volume is different by a factor of ten – or in other words, the directionality of $A_1$'s and $A_2$'s vectors is identical, but their lengths are different:

|       | RT $B_1$ | @$B_2$ |
|-------|----------|--------|
| $A_1$ | 50       | 20     |
| $A_2$ | 5        | 2      |

The various available measures of vector comparison can be affected by such differences. This may be welcome in contexts where activity volumes matter; elsewhere, however, we would still want to treat two accounts with the same interaction patterns as equivalent even if one account was somewhat more active than the other. Cosine similarity chiefly takes into account the angle between two vectors, rather than their lengths, and is therefore relatively robust; other similarity measures, such as Euclidean distance, are affected by differences in length considerably more strongly. Alternatively, to avoid impacts from different vector lengths, we can normalise the values for each dimension of an account's vector by dividing them by their sum (in our example above, 50+20=70 for $A_1$ and 5+2=7 for $A_2$:

|       | RT $B_1$ | @$B_2$ | Total      | *Normalised RT $B_1$* | *Normalised @$B_2$* |
|-------|----------|--------|------------|-----------------------|---------------------|
| $A_1$ | 50       | 20     | 50+20 = 70 | 50/70 = 0.71          | 20/70 = 0.29        |
| $A_2$ | 5        | 2      | 5+7 = 7    | 5/7 = 0.71            | 2/7 = 0.29          |

If we now apply cosine similarity to these *normalised* interaction vectors for $A_1$ and $A_2$, their practices will be regarded as identical – and in many applications of the practice mapping approach, this is likely to be the preferable outcome.

In spite of the eradication of activity volume differences that results from the normalisation of vector lengths, it will usually still remain useful to filter the least active accounts from the dataset before the systematic comparisons between the accounts' interaction vectors that practice mapping requires are applied. This has conceptual as well as practical reasons: at a conceptual level, the practices of accounts that show only very limited activity may well be similar to those of accounts that are highly active (for instance, after normalising, the practices of an account $A_1$ that @mentions each of $B_1$ and $B_2$ once can be considered to be identical to those of an account $A_2$ that @mentions each of $B_1$ and $B_2$ one thousand times), but for a highly active account we can safely assume that such a pattern represents clear intention, while for a largely inactive account the pattern may be merely accidental. To include such accidental matches in an analysis of the predominant practices that may be found in the dataset would be misleading; much as in many conventional network visualisations, then, it remains advisable in practice mapping, too, to define an appropriate threshold of minimum participation that accounts need to meet before they can be considered in the analysis. (The level of that threshold will depend on the nature of the dataset and the activities it describes; there is no universal solution for its selection.)

At a practical level, too, this reduction of the source dataset to those accounts which meet a minimum activity threshold is useful, as it reduces the computational load of the vector comparisons we must perform. Although, as we have noted, these cosine similarity calculations are symmetrical (so that we do not need to compare $A_1$ to $A_2$ *and* $A_2$ to $A_1$), for any given set of accounts $A_1$ to $A_n$ we still have to calculate the cosine similarity for each pair $A_i$ and $A_j$ where *j* is greater than *i*; this means that the computational load still grows at the rate of $n^2/2$. Removing accounts with low levels of activity from the set beforehand can thus substantially reduce this load.

The systematic calculation of the cosine similarities for the normalised activity patterns of each pair of accounts that remain in our dataset after filtering out the least active accounts will thus finally result in a similarity value between a pair of accounts $A_i$ and $A_j$ that ranges between 0 (entirely dissimilar) and 1 (completely identical). As noted above, we can now treat this as the representation of an *undirected* network between these accounts, and apply the standard tools of network analysis to this similarity network. Note that this similarity network is substantially different from conventional interaction networks, however, in that there is an edge between *every* pair of nodes, even if the weight of many such edges (i.e. the similarity between their practice patterns) will in many cases be close to or even exactly zero – in other words, even a comparatively small set of 1,000 accounts will have $1,000^2/2 = 500,000$ edges describing their pairwise similarities.

Since our practice mapping approach is predominantly interested in identifying groups of accounts with highly similar practices, it would therefore be appropriate to apply a further filter to the vector similarities network before attempting to analyse or visualise it: by setting a minimum threshold for cosine similarity values, we can filter out any edges that predominantly represent *dissimilarity*. Indeed, by retaining only those edges that point to considerable similarities in interaction patterns between accounts, it becomes even easier to identify those groups of accounts in our dataset that share common practices in their engagement with others.

## Practice Mapping and Its Interpretation

We demonstrate the results of the application of our practice mapping approach with a real-life example in fig. 5. Here, fig. 5a (left) shows a conventional network map, with the network comprised of directed retweet and @mention interactions between a set of accounts discussing a controversial tax debt recovery scheme, dubbed 'Robodebt', using the hashtag #robodebt in Australia between late 2016 and mid-2023 (cf. Dehghan & Bruns, 2023). We included only accounts with a combined total of at least 100 retweets and @mentions, and visualised the network using the Force Atlas 2 algorithm in the network analysis software Gephi (Jacomy et al., 2014). The overall structure of this network remains largely amorphous, as a result of the various limitations of conventional network analysis and visualisation that we have identified above. Meanwhile, to generate fig. 5b we drew on the same interaction data, generated normalised interaction vectors for all accounts that posted a combined total of at least 100 retweets and @mentions, calculated the pairwise similarities between these vectors, and filtered the edges represented by those similarities so that we retained only those edges with a cosine similarity value of at least 0.6. Finally, we visualised the undirected similarity network created by those edges using the Force Atlas 2 algorithm as implemented in Gephi, and used the Louvain modularity detection algorithm (Blondel et al., 2008) with a modularity threshold of 1.0 to detect and colour clusters in the network. (Simplified Google BigQuery SQL code for producing these network data from the source Twitter data is included in the Appendix.)

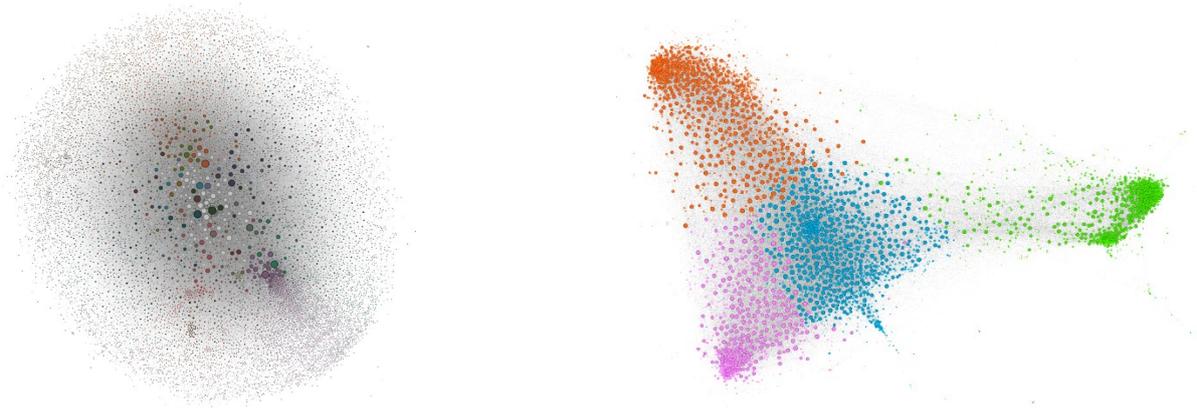

*Fig. 5a/b: Conventional interaction map (@mentions and retweets) for the #robodebt hashtag (left); practice mapping of account interaction similarities in the #robodebt hashtag (right). Nodes coloured based on Louvain modularity algorithm clusters. Any colour similarities between both maps are accidental.*

The resulting *map of practices* in fig. 5b clearly shows considerably more defined clusters of accounts than the *map of interactions* in fig. 5a; even though the various unidirectional or reciprocated interactions between accounts through retweeting and @mentioning that fig. 5a represents produced a graph that pulled participating accounts together into a tightly bound network with limited structural differentiation (or in other words, what network analysts have come to call a 'furball'), it turns out that underneath this flurry of @mentions and retweets between participants there are some clearly distinct practices which are held in common by subsets of the overall population.

Having identified – both visually through the practice mapping network and computationally through the modularity detection applied to that network – that these distinct practices are present, it is now also possible to determine what they represent. Approaches to doing so are necessarily dependent on the specific research context of the datasets being analysed, and may range from qualitative close reading of account details and post content to further quantitative analysis of the various practice clusters and their activities; here we can outline only a selection of standard approaches that may produce useful insights, and show how they helped us understand the divergent practices in our Robodebt example dataset.

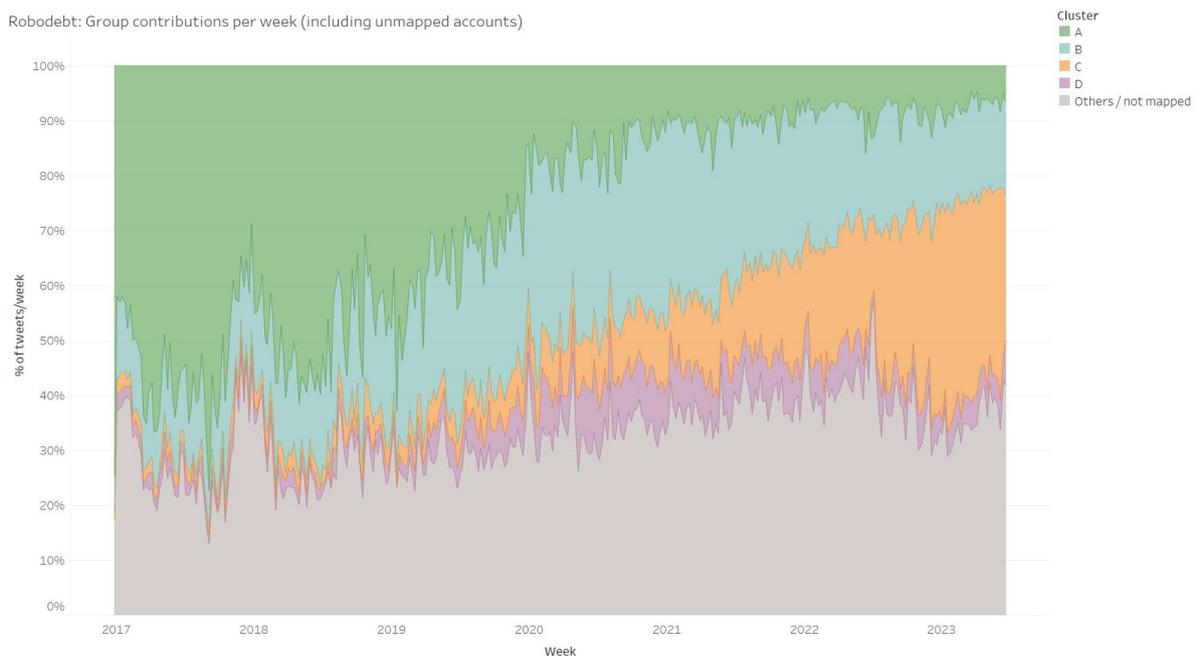

*Fig. 6: Contributions per week to the Robodebt debate by the four practice clusters identified in fig. 5b, and by all remaining accounts in the dataset*

First, we examine the contribution of the four practice clusters identified by our practice mapping in fig. 5b to the overall volume of activity in our dataset. We pay particular attention here to their contribution over time, since the Robodebt dataset covers a period of more than six years and we speculated that account participation practices may have evolved over time. Fig. 6 shows the relative contribution of each of these four clusters, as well as of any other accounts in our Robodebt dataset that were not included in our practice mapping analysis because they did not meet the required threshold of a combined total of at least 100 retweets and/or @mentions; it retains the same cluster colours as shown in fig. 5b. It is clearly evident that the group of accounts in cluster A dominate the early stages of the Robodebt debate, at times contributing more than 50% of all weekly tweets, and that accounts in groups B, C, and D increase their posting activity only at various later stages (with B dominating the mid-period and C the late stages of the timeframe covered here, and D making a minor contribution especially between 2019 and 2022).

Second, we separately examine each cluster of the practice mapping network in turn, and for that cluster calculate the total weighted degree of each of its accounts (note that as the practice mapping network is undirected, there can be no distinction between weighted indegree and weighted outdegree here). The weighted degree measure is the sum total of the weight of all edges connecting an account to the rest of the network, and thus in a practice mapping network represents the sum of all similarity comparison values. Calculated only for the practice cluster that the account belongs to, weighted degree therefore helps us determine which accounts in the cluster are *most similar* to all other accounts in the same cluster. These accounts thus model the common practices that define a cluster in the practice map: their practices can be regarded as archetypical for the cluster as a whole. We suggest this approach rather than merely taking the sum or average of the practice patterns exhibited by all the accounts assigned to a given cluster because it is in the very nature of the practice similarity comparison framework we have outlined here that practice clusters consist of a central core of highly archetypical representatives surrounded by a periphery of less typical members; to interpret the common factors that make the cluster a cluster (and potentially a community) it is therefore most useful to examine its most typical representatives.

It may be possible to draw conclusions about the nature of each cluster already from the identities of these most typical accounts in the cluster; however, the practice mapping approach we have outlined here – at least if it chiefly takes into account active participation, e.g. through @mentions and retweets – tends to identify the most active accounts as central members of each cluster, rather than those that are being addressed the most. In a third step, therefore, it might be useful for instance to examine whom these core members @mention or retweet most often, but perhaps also which hashtags they use most frequently, or what domains they share most regularly. This is likely to require some qualitative interpretation, and practice mapping is thus very clearly a mixed-methods rather than purely quantitative approach: while as part of the examination of these distinct practices we can quantitatively determine the most common interaction targets for the accounts representing each practice cluster, who or what these interaction targets represent (at a cultural, social, or political level) remains a qualitative question.

Combining these two steps for the Robodebt example, we first selected the ten most central accounts in each of the four key clusters identified in fig. 5b; we then identified the ten most frequent targets of these core representatives' @mentions and retweets, respectively. This process produced some substantially different patterns of interaction and attention for each of the four clusters, and combined with the temporal patterns in fig. 6 as well as our further background knowledge about the Robodebt case enables us to formulate a preliminary interpretation of these clusters and their divergent practices:

- Cluster A (green) represents early activists opposing the Robodebt scheme. They predominantly retweet their own community leaders, including a collective account (@not_my_debt) set up to protest the scheme, and @mention Australian government and opposition leaders as well as the Centrelink agency administering the scheme to raise awareness of its problems and demand action to remedy them.

- Cluster B (blue), prominent especially during 2019 and 2022, represents the politicisation of the Robodebt issue for party-political purposes. Members frequently @mention the then-Prime Minister Scott Morrison, as well as other ministers of the conservative government, but prominently retweet only Labor party and union movement accounts as well as the Twitter accounts of ordinary left-wing users. Notably, activity in this cluster stagnates following the change of government to the Labor party in 2022.
- Cluster C (orange) is mostly active from 2022 onwards, and especially increases its posting volume from August 2022, as the Royal Commission into the Robodebt Scheme begins its public hearings. Its practices are somewhat similar to those of cluster B, yet feature somewhat more mentions of media accounts (and especially those related to the national public broadcaster, the ABC) – most likely because its members are referring to or critiquing the ABC's live coverage of the Commission's hearings.
- Cluster D (purple) broadly follows the diachronic activity patterns of cluster B, if at a much lower volume (see fig. 6); fig. 5b, too, shows it to be closely aligned with that larger cluster. Our preliminary analysis appears to indicate a somewhat stronger alignment with the Australian union movement rather than the Labor Party, and this may drive the small distinctions between these two clusters. Further analysis will need to confirm this assumption.

We present these preliminary interpretations here not as a complete analysis of the distinctive participation practices in the Robodebt example, but in order to illustrate the analytical insights that the practice mapping approach can contribute. We note here, for example, that the brief practice mapping demonstration we have conducted is based solely on interaction data (@mentions and retweets), without taking into account any other of the potential attributes we could also have utilised to further describe each account's distinct practice patterns (e.g. hashtags used, URLs shared, periods of activity, profile information, etc.) – and yet it has already managed to identify several groups of accounts that are distinct not only in their interaction practices, but also in *when* over the course of the total period covered by the dataset they were especially prominent. The addition of further account activity attributes to the practice vectors for each account would likely enable an even sharper picture of distinct practices to emerge, but this brief demonstration serves as a sufficient indication of the practice mapping framework's utility.

In identifying these distinct practice clusters, we ought also to confront the question of whether these *clusters* can be said to represent genuine *communities* of practice (Lave & Wenger, 1991). As we have noted above, even though modularity detection approaches such as the Louvain algorithm we used in figs. 5a and b are often described as 'community detection' tools, in the first place they can only identify network clusters; whether such groups of accounts – in the case of social media networks – are also genuine communities in a social science sense will depend on whether their participants are mutually aware of each other, and indeed consider themselves to be members of a community with shared beliefs, values, and ideas (cf. Baym, 1998). The study of such self-identification is likely to require further data beyond what is available in any given social media dataset (e.g. through interviews or other ethnographic work), but the question of mutual awareness can be addressed at least in part through the data we have.

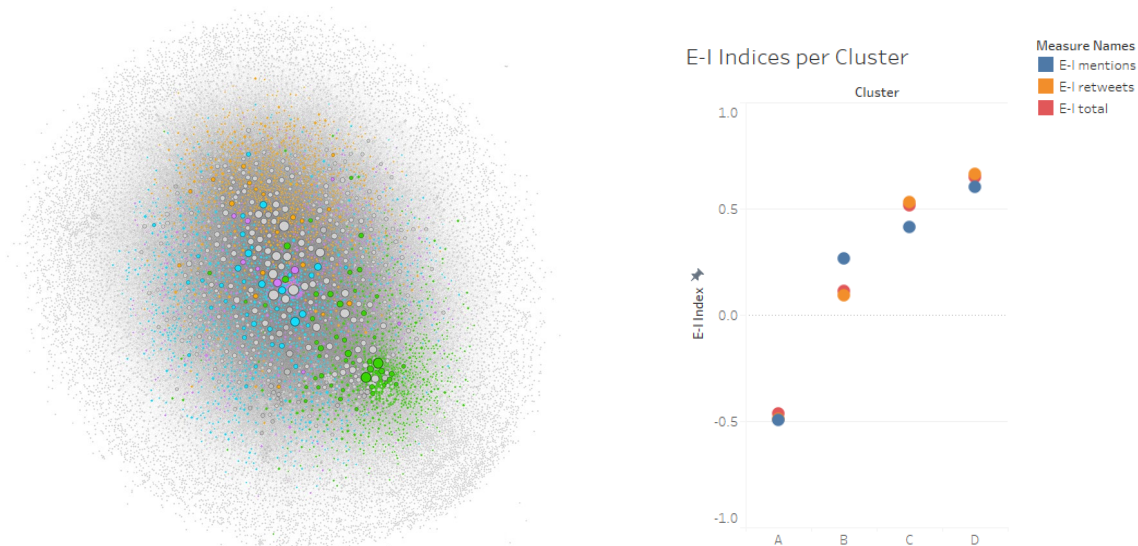

*Fig. 7a/b: a) Original 'furball' network map, with accounts belonging to the four major practice clusters highlighted in each cluster's colour; b) E-I Index values for each of the four key clusters, divided by tweet types: @mentions (blue), retweets (orange), @mentions and retweets (red)*

Several approaches may produce useful insights into this question, and two of these are illustrated by fig. 7. First, taking the example depicted in fig. 5, we transferred the colouring of clusters in the practice network (fig. 5b) back to the original interaction network (fig. 5a), to examine visually where in that network our practice clusters are located. The result of that process is shown in fig. 7a. Here, we can see that even amidst the generally shapeless structure of the original interactions network, the accounts belonging to distinct practice clusters are unevenly distributed across the overall map. In light of the limitations of conventional network mapping that we have outlined, it is unsurprising that the practice clusters are not neatly arranged in distinct regions of the interaction network – otherwise there would be no need for the practice mapping approach in the first place –, but we can nonetheless see some concentrations of cluster members in specific parts of the interaction network. While the other practice clusters intermingle to a considerable degree, this is especially pronounced for the original activist accounts contained in Cluster A (in green), whom the practice map in fig. 5b also showed to be particularly distinct. This points to the fact that – alongside other accounts that may be drawn into these regions because they are frequently engaged with by these practice cluster members, but do not actively tweet into the #robodebt hashtag – there is some degree of preferential interaction between practice cluster members that would suggest mutual awareness, especially in Cluster A.

Further, it can also be instructive to calculate additional metrics that compare the engagement of practice cluster members amongst themselves and with other parts of the network. Of particular use here is the E-I Index (Krackhardt & Stern, 1988), which can be used to compute a value from +1 to -1 that describes the relative external or internal focus of a group's interactions with others:

$$E\text{-}I\ Index = \frac{\#\ External\ Interactions\ -\ \#\ Internal\ interactions}{\#\ External\ Interactions\ +\ \#\ Internal\ Interactions}$$

For any given practice cluster, a value of +1 for its combined interactions would indicate that members of the cluster engage *only* with outsiders, and *never* amongst themselves (implying no mutual awareness whatsoever), and a value of -1 would mean that members engage *only* amongst themselves, and *never* with outsiders (implying a strong awareness of and preference for in-group communication only). Importantly, the E-I Index can also be computed separately for retweets and @mentions, of course, and this may be useful where these and other communicative affordances are used in distinctly different ways (e.g. using retweets to amplify in-group views and using @mentions to attack outsiders). Fig. 7b, then, presents the E-I Index values for each of the four practice clusters in fig. 5b, and again shows considerable differences between the clusters. Cluster A in particular is distinguished by its considerable inward focus: for all tweet types, it scores E-I Index values close to -0.5, indicating a strong inward focus. This aligns with our interpretation of this first cluster as representing early activists organising and raising awareness about the Robodebt scandal, not least also by coordinating amongst themselves through @mentions and amplifying each other through retweets. Subsequent clusters, by contrast, are progressively more outward-focussed: while Cluster B shows relatively neutral E-I Index values for retweeting and (less so) for @mentions, Clusters C and especially D score E-I Index values near or above +0.5 for their tweets of various types. This accords with our interpretation of these clusters as later arrivals: coming to the topic later, these users have an increasingly large number of accounts from outside their own groups to choose from as they @mention and retweet participants in the Robodebt debate.

Setting strict requirements for mutual awareness and a preference for in-group engagement, then, we might consider only Cluster A to represent a true online community: its members show a clear and significant preference for engaging in the first place with each other, rather than with outsiders. With the benefit of hindsight, we can confirm this interpretation: a key actor in this early activist cluster has reported that they built a strong community of like-minded activists in order to perform their Robodebt-related activism (Topsfield, 2023). However, as the overall interaction network in figs. 5a and 7a shows, there also is considerable interaction between accounts throughout the entire Robodebt dataset; a more balanced reading of our findings here, therefore, might be that Cluster A, not least by virtue of including the earliest and most committed activists, represents the central core of the *overall* Robodebt community, and that the later-arriving Clusters B, C, and D constitute subsequent layers of that community that accumulated around that central core – and, at least in the case of Cluster B, also established an alternative, if less distinct, second centre which (in our reading) focussed more on the political implications of the scandal than the flaws of the Robodebt scheme itself.

These two simple analyses are far from exhaustive, however: practice clusters may also be examined in a variety of other ways. In particular, it is possible that accounts with highly similar posting patterns may be coordinating their activities; it may therefore be useful to apply key tools like CooRNet (Giglietto et al., 2020) or the Coordination Network Toolkit (Graham et al., 2024) to the posts made by the members of each practice cluster in order to detect the presence of such (authentic or inauthentic) coordination. Whether such coordination is authentic or artificial, however, remains a matter for interpretation – prosocial activists may very legitimately coordinate their posting patterns to ensure greater visibility for their cause, for instance, while state-sponsored influence operations might engage in similar coordination for considerably more nefarious purposes.

## Mapping Practices beyond Retweets and @mentions

We have so far focussed predominantly on the mapping of shared interaction practices, using retweets and @mentions on Twitter as a convenient, well-known example. The application of the practice mapping approach to these kinds of interactions directs our focus in the first place to the practices shared between accounts that actively contributed posts to the dataset; contrary to conventional network maps, at whose centres we usually find the accounts of major news organisations, politicians, and other key stakeholders who are @mentioned and/or retweeted by all sides of a given issue, the map of practice clusters tends instead to highlight groups of those accounts that contributed most actively and (in terms of their interaction patterns) most consistently to the debate. (Usefully, this tends to exclude spam accounts, whose posting patterns are usually too idiosyncratic to be similar to genuine participants' practices.) As we have shown, the key targets of these clusters' interactions emerge only in a second step where we interpret the clusters that the practice mapping has identified.

However, in addition to the practice vectors describing the outgoing interactions from each account, we could just as easily also construct and compare a second set of vectors describing the incoming interactions received by each account. We can do so in an entirely separate analysis – i.e. considering and comparing *only* those incoming interaction vectors – or combine incoming and outgoing interactions into a single vector, to which we then apply our cosine similarity comparison (in the latter case, the outgoing and incoming components of the vector should be normalised separately, using the total sum of the account's sent and received interactions, respectively).

The former approach – focussing *only* on incoming interactions – is a form of practice mapping in the sense that it can be used to identify clusters of accounts that are frequently *targetted* by the same posters and their practices: this approach might identify, for instance, a group of political accounts that are commonly retweeted by their supporters and @mentioned by their opponents. The latter produces a combined map of similarities in both the accounts' own practices and those of other accounts targetting them; this combination is likely to water down any clear distinctions in either form of practices if those outgoing and incoming practices do not clearly overlap with each other; however, in some cases it may also produce further insights. For instance, if a group of accounts is targetted in distinct ways by the various groups on both sides of a political debate, but chooses to engage only with one of those sides – practicing what Dehghan has called "active passivity" towards the other side (2020, p. 228) – this interactive choice should stand out clearly.

We have used examples of the communicative affordances provided by Twitter in the discussion so far, since most scholars in our field will be familiar with these forms of interaction; however, subject to data availability it is just as easy to construct and compare practice vectors that represent an account's interactions on other social media platforms. Retweets and @mentions on Twitter are equivalent to boosts and @mentions on Mastodon, for instance; on other, more forum-based platforms, we might distinguish between comments, replies, and even up- and downvotes.

However, practice mapping does not need to limit itself to *interaction* practices in this explicit sense. In much the same way that we have constructed practice vectors that represent an account's engagement with other accounts, we can also construct vectors that show the account's practices in using hashtags, embedding images or videos, sharing external content (at the domain or URL level), or addressing certain topics. Indeed, the range of vectors we might consider here is limited only by what attributes of the account's posts (or in fact any other information about the account) can be quantified: this also includes the results of any manual or computational coding efforts (e.g. of the account's posts expressing support or opposition for specific political questions, or of the account profile's self-identification in terms of location, interests, or political stance) and background information from other sources (e.g. the party affiliations of politicians' accounts). In each case, these vectors can again be normalised against the total volume of the account's posts or other relevant measures, in order to avoid interference from different underlying levels of platform activity for each account. For example, the normalised hashtag vectors for accounts $A_1$ and $A_2$, using hashtags #$H_1$ and #$H_2$ as well as posting some tweets without hashtags, might be calculated as follows:

|  | #$H_1$ | #$H_2$ | none | Total | *Normalised #$H_1$* | *Normalised #$H_2$* | *Normalised none* |
| --- | --- | --- | --- | --- | --- | --- | --- |
| $A_1$ | 30 | 20 | 50 | 100 | 30/100 = 0.3 | 20/100 = 0.2 | 50/100 = 0.5 |
| $A_2$ | 3 | 2 | 5 | 10 | 3/10 = 0.3 | 2/10 = 0.2 | 5/10 = 0.5 |

These extensions of interaction practice mapping to include other practice aspects are perhaps best demonstrated by the discursive practice of engaging with certain topics. Already, many topic modelling approaches identify a set of distinct topics from an overall corpus of data and then calculate the relative affinity of each unit of text with these various topics, expressed for instance as a value between 0 and 1. This approach could be applied separately to each individual post or, alternatively, to the combined total of all posts made by a given account, and would then show the account's relative affinity with each in the full list of identified topics. By converting these into a vector and comparing these topic vectors for each pair of accounts, we can thus compute and map the similarity of topic choices for all accounts, producing either a practice mapping that shows these topical similarities in isolation (as a specific aspect of their practices), or a mapping that combines this aspect of the accounts' practices with other aspects (such as their interactive practices, approaches to information sourcing, etc.).

In pursing such a combined practice mapping approach, then, we must consider how best to integrate the different practice aspects we have described through the vectors that we have constructed for each account. Several approaches are possible. First, we could construct separate vectors for each practice aspect, calculate the similarities between these various vectors for each pair of accounts, and add up the similarity assessment for each aspect to produce a combined similarity value. In doing so, it would also be possible to award different weightings to these similarity scores: for instance, we might consider a strong similarity in interactive practices to count for more than a strong similarity in hashtag practices, and a strong similarity in the use of external sources to count for even more than these. For example, given individual similarity values for a pair of accounts $A_1$ and $A_2$, then, their total similarity score would be calculated as:

$$\text{similarity}_{total}(A_1,A_2) = \alpha \times \text{similarity}_{interactions}(A_1,A_2) + \beta \times \text{similarity}_{hashtags}(A_1,A_2) + \gamma \times \text{similarity}_{sources}(A_1,A_2)$$

The specific values for α, β, and γ will need to be carefully selected and justified, of course. Once the total similarities are calculated, however, the same analytical and interpretive processes as we have described them above for a single interaction similarities network can be applied to the combined practice mapping outcome; the practice clusters identified from it will then be the result of shared practice patterns across the various component practices contributing to the total similarity assessment.

A second option, which constitutes a simplified version of this first approach, is to combine the various vector components directly into one composite vector for each account, and to calculate the cosine similarities for each pair of practice vectors as usual. This substantially reduces the number of cosine similarity comparisons that must be performed (again only one per pair of accounts, rather than one for each practice component and pair), and thereby also lessens the computational load required. It does remove the opportunity to weight the relative importance of the various practice components that contribute to the combined vector, however, which may remain desirable.

Beyond these simple approaches to the (weighted or unweighted) addition of practice similarities, however, vector mathematics (algebraic topology) offers several other approaches, too. For instance, mapping *practice manifolds* – bundles of vectors that engage in similar practices in the higher-dimensional vector space – as lower-dimensional spaces that embed and convey the vernacular of a practice, can be employed at scale. Such practice manifolds, similar to the nonlinear dimensionality reduction algorithm Uniform-Manifold Approximation Projection (UMAP; McInnes et al., 2018), are Reimann manifolds that use distance metrics rather than a $O(n^2)$ pairwise cosine similarity algorithm to compare vectors. This also addresses issues of computability and the Curse of Dimensionality one encounters when embedding practices at scale.

However the various practice components are combined in any given case, what results from this process is an output that combines these distinct components into a single practice map, enabling the identification of the clusters of accounts that share strongly similar practices, and the examination and interpretation of the particular shared practices around which these clusters have formed. Especially for complex communicative contexts where networks of interaction alone fail to produce conclusive insights into the uses of social media for public debate, we argue that this practice mapping approach can offer considerably richer understandings.

## Conclusion and Outlook

This article has introduced practice mapping as a novel and powerful new technique for the analysis of social media data; as we have shown, it is able to identify distinct patterns of activity where conventional network mapping fails to provide sufficient clarity. In demonstrating its utility, we have applied this approach to data from the now defunct Twitter API, since many scholars in our field will be familiar with the features of such datasets; however, it should be noted that practice mapping excels especially also in situations where data on the structures of interaction networks between participants are limited: this is the case for instance for platforms like Facebook, where sources such as CrowdTangle or the Meta Content Library offer information on the text, image, and URL content *of* public pages and groups, but available information on connections *between* such spaces covers only the relatively uncommon practice of content on-sharing from one Facebook space to another. By treating URL, image, and textual patterns as equally meaningful practice vectors alongside any observable on-sharing practices, our practice mapping approach makes it possible to draw connections between such public spaces on Facebook based on their shared practices, thereby substantially enhancing the insights we are able to generate from such data.

Indeed, while we have presented the practice mapping approach here chiefly as examining patterns of similarity between *individual* accounts, this application to public pages and groups already points to the fact that it is also possible to use this approach to study patterns of similarity between *collective* online spaces. It would be possible, for instance, to treat each individual subreddit on Reddit as an object of study: here, in addition to the overall textual and outlinking patterns observed in the subreddit, information on its participants and their respective volume of contributions could be converted into a further vector to describe the subreddit, and used to calculate a new similarity score that describes the similarity between two subreddits' account populations and their posting patterns, for instance. (It would also be possible to conduct separate practice mapping analyses both within and across subreddits, of course, or at the level both of individual accounts and collective subreddits, in order to arrive at a hierarchical mapping that combines multiple levels of specificity.)

Similarly, opportunities to generate additional data points from the available social media data are limited only by the researcher's imagination. Using manual coding, computational processing, or AI-supported classification, it is possible to generate and extract a wide range of further attributes for each individual account or communicative space, which may then be systematically compared across the entire population as part of the practice mapping process. Such additional attributes could include assessments of the veracity, sentiment, or toxicity of the content posted by an account, for instance, or code for the distinct entities named, narrative frames used, or argumentative claims made. We hope that our own future work as well as the extensions of this approach by other scholars will provide a wide range of methods for the extraction of such attributes, and for their operationalisation in the practice mapping process.

At this attribute extraction as well as at the network interpretation stage of the process, it is especially evident that practice mapping is a thoroughly mixed-methods approach. While it involves a central computational component that facilitates the systematic comparison of available practice attributes across the entire population of accounts and produces a network map based on their similarities, we stress emphatically that the aim of this approach is not to simplistically quantify complex interactions between human and non-human participants in a communicative environment. Rather, practice mapping both enhances the systematic analysis and comparison of qualitative attributes across a large population of participants, and provides more rigorous foundations for the identification and interpretation of common patterns in their collective practices. Where the quantitative and qualitative analysis of these common patterns supports such an interpretation, it enables us, finally, to state with significantly greater certainty than previously that we have identified *communities of practice*.

## Acknowledgments

This research was funded by the Australian Research Council through the Australian Laureate Fellowship project *Determining the Drivers and Dynamics of Partisanship and Polarisation in Online Public Debate* and the Australian Future Fellowship *Understanding Intermedia Information Flows in the Australian Online Public Sphere*.

## Appendix

In this appendix we provide two brief queries that convert conventional social media data into a practice map. As in the article proper, we use Twitter data as an example here, and assume that such data are stored in the popular cloud data service Google BigQuery or a storage service with similar SQL functionality; the query below should also be easy enough to translate into Python, R, or other popular data processing languages, however. While actual data formats will vary depending on the tool used to collect the data, we assume that at a minimum, researchers have or able to create a table that contains data in the following form:

| tweet_id | author_id | target_id | tweet_type |
|---|---|---|---|
| 12345 | 111222 | 333444 | retweet |
| 12345 | 111222 | 666777 | mention |
| 34567 | 333444 | 666777 | mention |
| 56789 | 111222 | 555444 | mention |

Here, **tweet_id** is the numerical ID of each tweet, **author_id** is the account ID of the tweet's author, **target_id** is the account ID of any accounts addressed by the tweet (e.g. retweeted, @mentioned, quoted, etc.), and **tweet_type** describes the type of interaction with the target account. Note that a single **tweet_id** may address multiple targets in various different ways: for instance, the tweet may contain a quote tweet and several @mentions, and may itself be retweeting an earlier post.

We use the following BigQuery SQL query to convert this interactions table into a table of interaction vectors:

```sql
SELECT
  author_id,
  ARRAY_AGG(STRUCT(interaction AS interaction,
    tweet_count AS tweet_count)) AS interactions,
  ARRAY_AGG(STRUCT(interaction,
    tweet_count/total_tweet_count AS normalised_tweet_count)) AS normalised_interactions,
```

```sql
    MAX(total_tweet_count) AS total_tweet_count
FROM (
  SELECT
    interactions.author_id,
    interactions.interaction,
    interactions.tweet_count,
    total_count.total_tweet_count
  FROM (
    SELECT
      author_id,
      interaction,
      COUNT(DISTINCT tweet_id) AS tweet_count
    FROM (
      SELECT
        tweet_id,
        author_id,
        CONCAT(target_id, " ", tweet_type) AS interaction
      FROM
        `[project].[dataset].interactions`
    )
    GROUP BY
      author_id,
      interaction
    ) AS interactions
  LEFT JOIN (
    SELECT
      author_id,
      COUNT(DISTINCT tweet_id) AS total_tweet_count
    FROM
      `[project].[dataset].interactions`
    GROUP BY
      author_id
    ) AS total_count
  ON
    total_count.author_id = interactions.author_id
  )
GROUP BY
  author_id
```

This produces the following data structure, which we store in a new **interaction_vectors** table in BigQuery:

| author_id | interactions | | normalised_interactions | | total_tweet_count |
|---|---|---|---|---|---|
| 111222 | 333444 retweet | 90 | 333444 retweet | 0.9 | 100 |
| | 666777 mention | 10 | 666777 mention | 0.1 | |
| 333444 | 666777 mention | 30 | 666777 mention | 0.6 | 50 |
| | 666777 retweet | 20 | 666777 retweet | 0.4 | |

Here, there is one row for each **author_id**, which contains the n-dimensional vectors **interactions** and **normalised_interactions** and a **total_tweet_count** for the account.

Using the BigQuery function ML.DISTANCE, which calculates the distance between two vectors (i.e. the inverse of similarity), we then generate the practice mapping data which can be imported into Gephi or another network analysis tool for visualisation. In the following query, we rename the output columns to **Source**, **Target**, and **Weight** to anticipate the column names Gephi requires when importing data. In order to reduce the computational load of the pairwise comparison between accounts, we filter the **interaction_vectors** table for accounts with a minimum of 100 interaction tweets (the threshold used in the article proper). Further, we filter the output for a minimum cosine similarity of 0.6, in order to remove those edges from the network which predominantly indicate *dissimilarity* rather than similarity. (Such threshold values should be adjusted to the specific context and aims of each study, of course.)

```sql
WITH
  Vectors AS (
  SELECT
    author_id AS id,
    normalised_interactions AS vector
  FROM
    `[project].[dataset].interaction_vectors `
  WHERE
    total_tweet_count >= 100
  )

SELECT
  *
FROM (
  SELECT
    SourceId AS Source,
    TargetId AS Target,
    1 - ML.DISTANCE(SourceVector, TargetVector, 'COSINE') AS Weight
  FROM (
    SELECT
      a.id AS SourceId,
      a.vector AS SourceVector,
      b.id AS TargetId,
      b.vector AS TargetVector
    FROM
      Vectors AS a
    CROSS JOIN
      Vectors AS b
    WHERE
      a.id > b.id
    )
  )
WHERE
  Weight >= 0.6
```

This, then, finally produces a new table in the form **Source, Target, Weight**, which can be exported as a comma- or tab-separated file and imported into Gephi, using its Data Laboratory functionality. We stress here again that it is critically important to import this network dataset into Gephi as an *undirected* network, as the cosine similarity values contained in the **Weight** column represent mutual edges between accounts.